\documentstyle[prl,aps]{revtex}

\def\de{{\delta}}

\def\grad{{\nabla}}

\def\bg{{\bf g}}

\def\bbeta{{\mbox{\boldmath $\beta$}}}
\def\bpi{{\mbox{\boldmath $\pi$}}}

\def\cC{{\mathcal{C}}}
\def\cR{{\mathcal{R}}}

\def\d{{\partial}}
\def\dzeroh{{\hat\partial_0}}
\def\tdzero{{\tilde\partial_0}}

\def\Lie{{\pounds}}

\def\Ham{{\cal H}}
\def\tHam{{\tilde \Ham}}

\def\cX{{\mathcal{X}}}
\def\bR{{\bar R}}

\def\bgrad{{\bar\nabla}}

\newcommand\beq{\begin{equation}}
\newcommand\eeq{\end{equation}}
\newcommand\bea{\begin{eqnarray}}
\newcommand\eea{\end{eqnarray}}

\begin{document}

\draft
\twocolumn[\hsize\textwidth\columnwidth\hsize\csname
@twocolumnfalse\endcsname

\title{Hamiltonian Time Evolution for General Relativity}

\author{Arlen Anderson and James W. York, Jr.}
\address{
Dept. Physics and Astronomy,
Univ. North Carolina,
Chapel Hill NC 27599-3255}
\date{March 18, 1998}

\maketitle
\tightenlines
\vspace{-3cm}
\hfill IFP-UNC-524

\hfill gr-qc/9807041
\vspace{2cm}
\
\begin{abstract}
\widetext
Hamiltonian time evolution in terms of an explicit parameter time is 
derived for general relativity, even when the constraints are not 
satisfied, from the Arnowitt-Deser-Misner-Teitelboim-Ashtekar action 
in which the slicing density $\alpha(x,t)$ is freely specified while 
the lapse $N=\alpha g^{1/2}$ is not.  The constraint ``algebra'' becomes 
a well-posed evolution system for the constraints; this system is the
twice-contracted Bianchi identity when $R_{ij}=0$. The Hamiltonian constraint 
is an initial value constraint which determines $g^{1/2}$ and 
hence $N$, given $\alpha$.  
\end{abstract}
\narrowtext
\vskip2pc]

A minor change\cite{Tei82,Ash87} in the Arnowitt-Deser-Misner (ADM) 
action principle \cite{ADM,Dir} 
leads to striking consequences for the understanding of general
relativity in Hamiltonian form.  Recent work on hyperbolic formulations
\cite{cbr83,FrR94,Fri96,cby95,aacby97b,acby97,cbya98} of general relativity 
indicates \cite{aacby97b,cbya98} that the freely specifiable quantity
that determines the slicing of spacetime is not the
lapse $N$ but the ``slicing density'' $\alpha(x,t) = N g^{-1/2}$
(where $g=\det g_{ij}$ is the determinant of the spatial metric).
Altering the action principle to take this into account leads to
a number of key results: 1) equations of motion that are equivalent to
$R_{ij}=0$, not $G_{ij}=0$, 2) a Hamiltonian vector field that generates
time evolution even when the constraints are not satisfied, 
3) a constraint algebra that is a 
homogeneous symmetric hyperbolic system that dynamically preserves 
the constraints, and 4) a new
understanding of the Dirac algebra. 

Before beginning, let us fix notation and note a few elementary results.  
We work on a manifold $\Sigma\times R$
with a ``foliation-adapted'' cobasis for the metric 
\beq
\label{metric}
ds^2 = - N^2 (dt)^2 + g_{ij} (dx^i + \beta^i dt) (dx^j + \beta^j dt).
\eeq
Here, $N$ is the lapse function (a space scalar) and $\beta^i$ is
the spatial shift vector. 
Quantities with overbars denote spatial quantities, in particular
$\bR$ the spatial curvature scalar obtained from $g_{ij}$ 
and $\bgrad_i$ the corresponding spatial
covariant derivative.  We also introduce the extrinsic
curvature $K_{ij}$ and its trace $H\equiv K^k\mathstrut_k$.  The
momentum conjugate to the metric is a density of weight one,
$\pi^{ij}=g^{1/2}(Hg^{ij} - K^{ij})$.  The natural time derivative
for evolution, $\dzeroh$, acts in the normal future direction to the 
spacelike slice and is
denoted by an over-dot.  It is given by $\dzeroh = \d_t - \Lie_\bbeta$,
where $\Lie_\bbeta$ is the Lie derivative along the shift $\bbeta$.

We have found in work on hyperbolic formulations of the equations of 
evolution of general relativity which have no unphysical characteristics
\cite{cby95,aacby97b,acby97,cbya98,aacby95,aacby96,cby97,aacby97a} that 
we must in essence use the Choquet-Bruhat
``algebraic gauge'' \cite{cbr83} to restrict the ordinary
lapse $N$. The weight-minus-one lapse (the ``slicing density'')
$\alpha = N g^{-1/2} = \alpha (x,t)$ is {\it freely} specifiable
while $N$ is not \cite{tmna}. (The slicing density $\alpha$ is also 
used prominently in \cite{Tei82} and \cite{Ash87,Ash88}.)
Indeed, if one computes $\dzeroh \log \alpha = f(x,t)$ from a given 
$\alpha(x,t)$, then one finds
\beq
\label{harmonic}
g^{1/2} \dzeroh \alpha = \dzeroh N + N^2 H = N f,
\eeq
the equation of harmonic time slicing \cite{cby95,aacby95}
with $f(x,t) = \dzeroh \log \alpha$ acting as a ``gauge source'' 
\cite{Fri85}.  Combining this with the 3+1 equation for $\dot H$ 
from the trace of (\ref{dK}) below, one obtains a quasi-linear
wave equation for $N$.  Every foliation is described by such a wave equation 
for some value of $\alpha$.  This wave equation for 
$N$ \cite{cby95,aacby97b,aacby97a} played a vital role
in the development of our first order symmetric
hyperbolic ``Einstein-Ricci'' system\cite{cby95,acby97,cbya98,aacby95}
and reflects the built-in {\it causality} which comes from working
with $\alpha$.   We conclude that $N$, which determines the proper time 
as $N\delta t$ between slices $t=t'$ and $t=t'+\delta t$, is 
a {\it dynamical variable} (cf. \cite{Ash88}), closely connected to 
$g^{1/2}$.  $N$ can also be seen to be determined 
from $\alpha$ and the Hamiltonian constraint,
written as the generalized and completed conformal ``Lichnerowicz
equation'' \cite{Yor79,Yor73,OMY74,cby83}; the explicit form
in \cite{OMY74} is the suitable one for present purposes.
{}From this perspective, the Hamiltonian constraint plays
its familiar role as an initial
value constraint which determines $g^{1/2}$ given a complete set of
freely specified data \cite{Yor73}.  
The important insight is that this in turn determines $N$ from $\alpha$,
so that the Hamiltonian constraint does not fix the time but
does fix the proper time $\tau$'s rate with respect to $t$: $d\tau/dt =
\alpha g^{1/2} =N$
along the normal $\partial_0$.

Motivated by these findings, we alter the undetermined
multiplier in the ADM canonical action principle from $N$ to $\alpha$.  
Using $\alpha$ has the affect of altering the Hamiltonian density 
from $\Ham$ to 
\beq
\tHam = g^{1/2} \Ham= \pi^{ij} \pi_{ij} - {1\over 2} \pi^2 - g\bR,
\eeq
the latter being of scalar weight plus two and a rational function of
the metric.  (Note we reserve the phrase ``Hamiltonian constraint''
to refer to the equation $\tHam=0$ and use ``Hamiltonian density''
for $\tHam$, which may not vanish, similarly for the momentum constraint
and density.)
This leads to Teitelboim's \cite{Tei82} and Ashtekar's \cite{Ash87} 
modification of the ADM action  ($16\pi G =1 = c$)
\beq
\label{ADMT}
S_{ADMTA}[\bg,\bpi;\alpha,\bbeta)= \int d^4 x\, (\pi^{ij}
\dot g_{ij} - \alpha \tHam),
\eeq
where we use Kucha\v r's notation indicating functional and explicit function
dependence and boundary terms are ignored here as they are not the
focus of the present analysis.  [There are no difficulties
in obtaining, for example, the ADM energy surface integral.  One 
requires the same asymptotic conditions as always, maintaining
$N\rightarrow 1 + O(r^{-1})$---not 
$\alpha \rightarrow 1 + O (r^{-1})$---while recalling $N=g^{1/2}\alpha$.]  
The vacuum case is 
considered, but to add minimally coupled matter and/or a cosmological
constant is straightforward.
The momentum density $\Ham_i$ has been absorbed into $\pi^{ij} \dot g_{ij}$
through use of the time derivative $\dzeroh$.  Explicitly, the
Lie derivative term in $\dot \pi^{ij}$ is, up to a divergence, 
\beq
2 \beta^i \bgrad_j \pi^{j}\mathstrut_i = - \beta^i \Ham_i.
\eeq

Consider a general variation of the modified Hamiltonian density
\bea
\label{tHamvar}
\de \tHam &=& (2 \pi_{ij} - g_{ij} \pi) \de \pi^{ij} +
(2 \pi^{ik} \pi^j\mathstrut_k - \pi \pi^{ij} \\
&& \hspace{-1cm}+ g \bR^{ij} -g g^{ij} \bR ) 
\de g_{ij} 
 - g (\bgrad^i \bgrad^j \de g_{ij} - g^{ij} \bgrad^k \bgrad_k \de g_{ij}).
\nonumber
\eea
Note that (\ref{tHamvar}) does not involve either the Hamiltonian
or momentum densities while,
in contrast, the variation of the ADM Hamiltonian density 
$\delta \Ham = \delta ( g^{-1/2} \tHam)$ does contain 
a term proportional to the Hamiltonian density.

Requiring that $S_{ADMTA}$ be stationary under a variation with respect to
$\pi^{ij}$ gives the definition of the extrinsic curvature
\beq
\label{ourg}
\dot g_{ij} = \alpha {\de \tHam\over \de \pi^{ij}} 
= \alpha (2 \pi_{ij} - g_{ij} \pi) \equiv -2 N K_{ij}. 
\eeq
Requiring that it be stationary under a variation with respect to 
$g_{ij}$ gives the equation of motion
\bea
\label{ourpi}
\dot \pi^{ij} &=& -\alpha {\de \tHam \over \de g_{ij}} 
= -\alpha g (\bR^{ij} - g^{ij} \bR)
- \alpha (2 \pi^{ik} \pi^j\mathstrut_k - \pi \pi^{ij}) \nonumber \\
&& 
 +g  (\bgrad^i \bgrad^j \alpha - g^{ij} \bgrad^k \bgrad_k \alpha).
\eea

The slicing density $\alpha$ and the shift $\beta^i$ are not to be
varied.  Instead the constraints are imposed on 
initial data and are preserved dynamically as shown below.  This 
is not an already parametrized theory in the usual sense.

Consider the familiar 3+1 identities
\beq
\label{dg}
\dot g_{ij} \equiv -2 N K_{ij} ,
\eeq
\beq
\label{dK}
\dot K_{ij} \equiv N (\bR_{ij} - R_{ij} + H K_{ij} 
- K_{ik} K_j\mathstrut^k - N^{-1} \bgrad_i \bgrad_j N).
\eeq
Also, recall the formula for the derivative of the determinant
of the 3-metric, 
$g^{-1} \dot g = g^{ij} \dot g_{ij} = -2 N H$.
Now we pass to canonical variables.
Using (\ref{dg}), (\ref{dK}), the time derivative of $\pi^{ij}$ is
computed to be identically
\bea
\label{pidotR}
\dot \pi^{ij}
&\equiv& N g^{1/2} (\bR g^{ij} - \bR^{ij}) -  N g^{-1/2} ( 2 \pi^{ik}
\pi^j\mathstrut_k - \pi \pi^{ij}) \nonumber \\
&& \hspace{-0.8cm}
+ g^{1/2} (\bgrad^i \bgrad^j N - g^{ij} \bgrad^k \bgrad_k N)
+ N g^{1/2} [ \cR^{ij}  ],
\eea
where $\cR_{ij} \equiv R_{ij} - g_{ij} R^k\mathstrut_k$.

We see that the equations of motion (\ref{ourg}) and (\ref{ourpi}) 
derived from the action principle
are (\ref{dg}) and (\ref{pidotR}) 
when $R^{ij}-g^{ij} R^k\mathstrut_k=0$.  Thus,
to say that (\ref{ourpi}) holds is to assert that $R^{ij}=0$.
The equations of motion hold strongly, independent of whether the
constraints are satisfied. This is not
true in the ADM formulation because of the presence of the Hamiltonian
density in their equation of motion for $\pi^{ij}$.

This difference can be explained more fully as follows.
{}From the definition of the Einstein tensor in terms of the Ricci tensor,
$G_{\mu\nu} \equiv R_{\mu\nu} - {1\over 2} g_{\mu\nu}
R^\sigma\mathstrut_\sigma$,
and the observation that $2G^0\mathstrut_0 \equiv
R^0\mathstrut_0 - R^k\mathstrut_k$, one derives the identity
\beq
\label{id2}
G_{ij} + g_{ij} G^0\mathstrut_0 \equiv R_{ij} - g_{ij} R^k\mathstrut_k.
\eeq
The vanishing of the right hand side does not depend on either the
Hamiltonian or momentum densities 
and is equivalent to $R_{ij} = 0$.  Clearly, it is also
equivalent to $G_{ij} = - g_{ij} G^0\mathstrut_0$.
Thus, while $R_{\mu\nu}=0$ and $G_{\mu\nu}=0$ are equivalent,
$R_{ij}=0$ and $G_{ij}=0$ are not equivalent as
equations of motion---unless the Hamiltonian density 
$\Ham = 2 g^{1/2} G^0\mathstrut_0$
vanishes exactly, that is unless the Hamiltonian constraint holds.  
The ADM action principle is equivalent to $G_{ij}=0$ and so also
requires $\Ham=0$ to be equivalent to (\ref{pidotR}). 
We recall that
the use of $R_{ij}$ has always been preferred by the French school,
pioneered by Lichnerowicz \cite{Lic44} and Choquet-Bruhat \cite{cb56}.

This raises an important principle: A constrained Hamiltonian theory 
should be well-behaved
even when the constraints are violated.  As discussed in \cite{aal98},
recent efforts
\cite{cby95,aacby97b,acby97,cbya98,aacby95,aacby96,cby97,aacby97a}
to achieve well-posed hyperbolic formulations of general
relativity, with only physical characteristics, can be understood in
this light as well.  From this point of view, the Hamiltonian and 
momentum densities are definite fixed combinations of the phase space 
variables, but their values may deviate from zero.
The form of the equations of motion should not
depend on these values.  When the constraints are satisfied
({\it i.e.}, the densities vanish), one is on the so-called constraint 
hypersurface, and
many unphysical degrees of freedom are frozen because relations among
many of the variables $(\bg,\bpi)$ are fixed.  When the constraints
are relaxed, the theory explores phase space away from the constraint
hypersurface.  The objective is to have a theory whose character
does not change dramatically when one moves off the constraint hypersurface.
Examples for which this is particularly relevant are
numerical applications where violation of the constraints is inevitable.
It seems that similar properties may be shared by the canonical
Ashtekar variables, but there are subtleties beyond
the scope of this paper that require
closer investigation (cf. \cite{Ash88}). 
[At this point, we should stress that the $R_{ij}=0$ equations given
by (\ref{ourg}), (\ref{ourpi}) or (\ref{dg}), (\ref{dK}) 
are not in themselves known to be well-posed, though they
have no unphysical characteristic speeds \cite{Fri96}.  They do,
however, lead to well-posed evolution of $\Ham$ and $\Ham_i$ as we shall
see.]

Introduce the smeared Hamiltonian 
\beq
\tHam_\alpha = \int d^3 x'\, \alpha(x',t) \tHam.
\eeq
The equation of motion for a
general functional on phase space $F[\bg,\bpi;x,t)$ is
\beq
\label{clPB}
\dzeroh F[\bg,\bpi;x,t) = - \{\tHam_\alpha, F[\bg,\bpi;x,t) \} +
\tdzero F[\bg,\bpi;x,t).
\eeq
Here, $\dzeroh$ is a total time derivative while $\tdzero$ is
a ``partial'' derivative of the form $\d_t - \Lie_\bbeta$ which only 
acts on explicit spacetime dependence.
The Poisson bracket is given by
$$
\{ F, G \} = \int d^3 x {\delta F \over \delta g_{ij}(x,t)}
{\delta G\over \delta \pi^{ij}(x,t)} -
{\delta F \over \delta \pi^{ij}(x,t)}
{\delta G\over \delta g_{ij}(x,t)}.
$$
It is evident that the equations of motion (\ref{ourg}), (\ref{ourpi})
are obtained by applying (\ref{clPB}) to the canonical variables
$(\bg,\bpi)$.  Also, observe that applying (\ref{clPB}) to $N=\alpha g^{1/2}$
produces (\ref{harmonic}). 

Time evolution is generated by the Hamiltonian vector field 
\bea
\cX_{\tHam_\alpha} &=& \int d^3 x \bigl\{  \alpha (2\pi_{ij} -
 \pi g_{ij}) {\de \ \ \over \de g_{ij}}
- [ \alpha g (\bR^{ij} - g^{ij} \bR) \\
&& \hspace{-1cm} +  \alpha ( 2 \pi^{ik} \pi^j\mathstrut_k - \pi \pi^{ij})
- g (\bgrad^i \bgrad^j \alpha - g^{ij} \bgrad^k \bgrad_k \alpha)
 ]{ \de \ \ \over \de\pi^{ij}} \bigr\}.
\nonumber
\eea
Again, this does not depend on the Hamiltonian or momentum densities, so it is
a good time evolution operator even away from the constraint 
hypersurface. (This observation was essentially made by
Ashtekar in footnote 17 of \cite{Ash87} but evolution was
mistakenly associated with $G_{ij}=0$).

By the product rule shared by $\dzeroh$ and the Poisson bracket, 
one computes the evolution equations for the constraints to be 
\bea
\label{eomHam}
\dot \tHam &=& -\{ \tHam_\alpha, \tHam \} = g \alpha g^{ij} \d_i \Ham_j +
2 g g^{ij} \Ham_i \bgrad_j \alpha, \\
\label{eomMom}
\dot \Ham_j &=& - \{ \tHam_\alpha, \Ham_j \} = \alpha \d_j \tHam +
2 \tHam \d_j \alpha,
\eea
where $\bgrad_j \alpha = \d_j \alpha + \alpha g^{-1/2} \d_i g^{1/2}$.
These equations correspond to (\ref{modBiC}) and (\ref{modBiCj})
below when $R_{ij}=0$. 
Thus, the Poisson brackets of the smeared Hamiltonian with the
unsmeared densities are seen to be well-posed evolution equations for
the densities. 

These equations can be shown to be equivalent to the twice-contracted 
Bianchi identities $\grad_\beta G^\beta\mathstrut_\alpha \equiv 0$
when $R_{ij}=0$ as follows.  Combine the identity (\ref{id2}) with the
twice-contracted Bianchi identities 
to obtain a transparent form of the Bianchi identities
\beq
\label{modBi1}
\grad_\beta G^\beta\mathstrut_0 \equiv \grad_0 G^0\mathstrut_0 +
\grad_j G^j\mathstrut_0 \equiv 0,
\eeq
\beq
\label{modBi2}
\grad_\beta G^\beta\mathstrut_j \equiv  \grad_0 G^0\mathstrut_j
-\grad_j G^0\mathstrut_0  + \grad_i [R^i\mathstrut_j - 
\delta^i\mathstrut_j R^k\mathstrut_k]
\equiv 0.
\eeq
(After the expression in square brackets is replaced using the 
vanishing of $R_{ij}$, these become equations
of motion rather than identities.)
Express the Bianchi identities (\ref{modBi1}),
(\ref{modBi2}) in 3+1 language,
using $\cC = g^{-1} \tHam=2 G^0\mathstrut_0$ and
$\cC_i = g^{-1/2} \Ham_i= 2 N G^0\mathstrut_i$.  A
calculation yields
\bea
\label{modBiC}
\dot \cC - N \bgrad^j \cC_j &\equiv& 2 \biggl( \cC_j \bgrad^j N + N H \cC
-N K^{ij} [\cR_{ij}] \biggr), \\
\label{modBiCj}
\dot \cC_j - N \bgrad_j \cC &\equiv& 2 \biggl( \cC \bgrad_j N
+{1\over 2} N H \cC_j  - \bgrad^i(N [\cR_{ij}]) \biggr),
\eea
where $\cR_{ij} \equiv R_{ij} - g_{ij} R^k\mathstrut_k$.
This system is clearly symmetric hyperbolic with only
the light cone as characteristic, and changing to $\tHam$,
$\Ham_i$ gives (\ref{eomHam}) and (\ref{eomMom}) when $R_{ij}=0$.
(Without considering our identities, or a Hamiltonian framework,
Frittelli\cite{Fri97} reached the same conclusion about well-posedness
of constraint propagation in the ``standard'' 3+1 formulation\cite{Yor79},
which uses $R_{ij}=0$, and its absence in the ADM equations,
with $G_{ij}=0$.
Related formulae were also obtained by Choquet-Bruhat and 
Noutchegueme \cite{CBN86}
for the evolution of matter sources
$\rho^{00}$, $\rho^{0i}$, where $\rho^{\beta\alpha} = T^{\beta\alpha}
-{1\over 2} g^{\beta\alpha} T^\mu\mathstrut_\mu$.)  

Return to the Hamiltonian formulation.
The Poisson bracket between two smeared Hamiltonians is
\bea
\label{PBHam}
\{ \tHam_{\alpha_1},\tHam_{\alpha_2} \} &=& -\int d^3 x\, g
(\alpha_1 \bgrad^i \alpha_2 - \alpha_2 \bgrad^i \alpha_1) \Ham_i \\
&=& -\int d^3 x\, g g^{ij}
(\alpha_1 \d_j \alpha_2 - \alpha_2 \d_j \alpha_1) \Ham_i \nonumber
\eea
This bracket expresses the consistency of
time evolution under different choices of $\alpha$.  The Jacobi
identity is
\bea
\label{Jacobi}
\{ \tHam_{\alpha_1}, \{ \tHam_{\alpha_2}, F \} \} -
\{ \tHam_{\alpha_2}, \{ \tHam_{\alpha_1}, F \} \} &=& \nonumber \\
&& \hspace{-1.5cm} = \{ \{ \tHam_{\alpha_1},  \tHam_{\alpha_2} \}, F  \}.
\eea
Because of the metric dependence in (\ref{PBHam}), one sees that the 
difference between evolution with $\alpha_2$
followed by $\alpha_1$ and the reverse is a spatial diffeomorphism
when $\Ham_i=0$ \cite{Tei73} (or when $\delta F/\delta \pi^{ij}=0$).

These results lead to a new understanding of the Dirac ``algebra'' of the 
constraints (cf. \cite{Tei73}).  As is well-known, the Dirac algebra is 
not the  spacetime diffeomorphism algebra.
The root of this is that the action (\ref{ADMT}) is invariant
under transformations generated by the constraints even when they are not 
satisfied \cite{Tei73b}. The equations which hold even when 
the constraints are not imposed are $R_{ij}=0$.  These equations
are preserved by spatial diffeomorphisms and
time translations along their flow, yet a general spacetime 
diffeomorphism applied to $R_{ij}=0$ mixes in the constraints. 
A comparison of (\ref{eomHam}), (\ref{eomMom}) and (\ref{modBiC}), 
(\ref{modBiCj})  shows the effect clearly.  The Bianchi identities 
are spacetime diffeomorphism invariant while the constraint evolution 
equations derived from the action principle are not. 
The equations (\ref{eomHam}), (\ref{eomMom}) and (\ref{modBiC}), 
(\ref{modBiCj}) differ precisely by terms proportional to $R_{ij}$.  

A second crucial understanding is the way in which 
the once-smeared form of the Dirac algebra (\ref{eomHam}), (\ref{eomMom})
ensures consistency of the constraints {\it via} a well-posed initial 
value problem.  If the constraints vanish initially, then they
always vanish in a corresponding physical domain of dependence. This 
dynamical mechanism for consistency follows from the dual role of $\tHam$ as 
part of the generator of time translations and
as an initial value constraint. 

It is worth re-emphasizing the altered role of
the Hamiltonian constraint.  The Hamiltonian constraint is an
initial value constraint from which $g^{1/2}$ is determined
as in the solution of the initial value problem \cite{Yor79}, which
then allows $N$ to be reconstructed from $\alpha$.
By virtue of (\ref{eomHam}), (\ref{eomMom}), once the initial
value problem is solved, it remains solved in a spacetime domain
dictated by causality.  The Hamiltonian
constraint does not express the dynamics of the theory; 
(\ref{clPB}) is the dynamical equation.  
The application of these ideas to canonical quantum gravity will
appear elsewhere \cite{AnYprep}.

This work was supported in part by National
Science Foundation grants PHY-9413207 and PHY 93-18152/ASC 93-18152
(ARPA supplemented).


\begin{thebibliography}{99}
\bibitem{tmna} Note that $N g^{-1/2}$ is called $\alpha^{-1}$ in
\cite{acby97}.
\end{thebibliography}
\end{document}